\documentstyle[12pt,epsfig,amssymb]{article}
\def\ap#1#2#3{Ann.\ Phys.\ (NY) #1 (19#3) #2}

\def\np#1#2#3{Nucl.\ Phys.\ B#1 (19#3) #2}
\def\pl#1#2#3{Phys.\ Lett.\ #1B (19#3) #2}

\def\re#1{(\ref{#1})}
\def\beq{\begin{equation}}
\def\eeq{\end{equation}}
\def\beeq{\begin{eqnarray}}
\def\beeqn{\begin{eqnarray*}}
\def\eeeq{\end{eqnarray}}
\def\eeeqn{\end{eqnarray*}}
\def\nome#1{{\label{#1}}}
\def\a{\alpha}
\def\b{\beta}
\def\g{\gamma}                  \def\G{\Gamma}
\def\de{\delta}                 \def\D{\Delta}
\def\eps{\varepsilon}

\def\l{\lambda}                 \def\L{\Lambda}
\def\m{\mu}
\def\n{\nu}
\def\ta{\tau}
\def\r{\rho}
\def\s{\sigma}                  \def\S{\Sigma}
\def\th{\theta}

\def\z{\zeta} 
\renewcommand{\AA}{{\cal A}}
\newcommand{\DD}{{\cal D}}
\newcommand{\SS}{{\cal S}}
\newcommand{\OO}{{\cal O}}
\newcommand{\WW}{{\cal W}}
\newcommand{\zzint}{\int d^4x\; d^2\th\,d^2\bt\;}
\newcommand{\cint}{\int d^4x\; d^2\th\;}
\newcommand{\acint}{\int d^4x\;d^2\bt\;}
\newcommand{\zint}{\int_{z}}
\newcommand{\spint}{\int_{p}}
\newcommand{\spqint}{\int_{pq}}
\newcommand{\sppint}{\int_{p'}}
\newcommand{\skint}{\int_{k}}
\newcommand{\sqint}{\int_{q}}

\newcommand{\lp}{\left(}
\newcommand{\rp}{\right)}
\renewcommand{\lq}{\left[}
\renewcommand{\rq}{\right]}
\newcommand{\lgr}{\left\{}
\newcommand{\rgr}{\right\}}
\newcommand{\identity}{1\hspace{-0.4em}1}
\newcommand{\no}{\nonumber}
\newcommand{\ph}{\phantom} 
\def\abs#1{\left|#1\right|}
\def\tr{\,\mbox{Tr}\,}
\def\frac#1#2{ {{#1} \over {#2} }}
\def\half{\mbox{\small $\frac{1}{2}$}}
\def\p{\partial}
\newcommand{\dpad}[2]{{\displaystyle{\frac{\partial #1}{\partial #2}}}}
\newcommand{\dfud}[2]{{\displaystyle{\frac{\delta #1}{\delta #2}}}}
\def\ie{\hbox{\it i.e.}{ }}      

\def\eg{\hbox{\it e.g.}{ }}      
\newcommand{\sde}{\de^8}
\newcommand{\ad}{{\dot\a}}  \newcommand{\bd}{{\dot\b}}  
\newcommand{\bD}{{\bar{D}}}
\newcommand{\bJ}{{\bar{J}}}

\def\cp{c_{+}}
\def\cm{c_{-}}
\def\bcp{\bar c_{+}}
\def\bcm{\bar c_{-}}
\def\bchi{\bar \chi}
\def\bxi{\bar\xi}

\def\bpsi{\bar \psi}
\def\bphi{\bar \phi}
\def\tgv{\tilde\g_{\scriptscriptstyle{V}}}
\def\gv{\g_{\scriptscriptstyle{V}}}
\newcommand{\bs}{{\bar\s}}
\newcommand{\bt}{{\bar\th}}
\newcommand{\smuaad}{\s^\m_{\a\ad}}
\def\ei{\varepsilon_{\footnotesize{i}}}
\def\ej{\varepsilon_{\footnotesize{j}}}
\def\ek{\varepsilon_{\footnotesize{k}}}
\def\el{\varepsilon_{\footnotesize{l}}}

\def\Wi{W^{\mbox{\scriptsize{int}}}}
\def\se{S_{\mbox{\footnotesize{eff}}}}
\def\st{S_{\mbox{\footnotesize{tot}}}}
\def\si{S_{\mbox{\scriptsize{int}}}}
\def\scl{S_{\mbox{\scriptsize{cl}}}}
\def\sgf{S_{\mbox{\scriptsize{gf}}}}
\def\sfp{S_{\mbox{\scriptsize{FP}}}}
\def\sbrs{S_{\mbox{\scriptsize{BRS}}}}
\def\sym{S_{\mbox{\scriptsize{SYM}}}}
\def\Pir{\Pi_{\mbox{\footnotesize{rel}}}}
\def\Piinv{\Pi_{\mbox{\footnotesize{inv}}}}
\def\Pit{\tilde\Pi}
\def\bG{\bar\Gamma}
\def\Gr{\G_{\mbox{\footnotesize{rel}}}}
\def\Gir{\G_{\mbox{\footnotesize{irr}}}}
\def\Gi{\G^{\mbox{\scriptsize{int}}}}
\def\Gio{\G^{\mbox{\scriptsize{int}}(0)}}
\def\De{\D_{\mbox{\footnotesize{eff}}}}
\def\DG{\D_{\G}}
\def\DGi{\D_{\G,\mbox{\footnotesize{irr}}}}
\def\DGb{\bar{\D}_{\G}}
\def\DGh{\hat{\D}_{\G}}
\def\DGhr{{\hat{\D}}_{\G,\mbox{\footnotesize{rel}}}}

\def\LdL{\L\partial_\L}
\def\UV{$\L_0\to\infty\;$}

\def\K{K_{\L\L_0}}
\def\Ki{K_{0\L}}
\def\Kiu{K_{0\L_{0}}}
\def\Kin{K_{\L_{0}\infty}}
\begin{document}
\begin{titlepage}

\begin{flushright}
UPRF-98-03 \\
February 1998
\end{flushright}
\vspace{.4in}
\begin{center}
{\large{\bf 
Wilson Renormalization Group
for Supersymmetric Gauge Theories
and Gauge Anomalies
}}
\bigskip \\ M. Bonini and F. Vian
\\
\vspace{\baselineskip}
{\small Universit\`a degli Studi di Parma \\ and\\
I.N.F.N., Gruppo collegato di Parma, 
\\viale delle Scienze, 
43100 Parma, Italy} \\
\mbox{} \\
\vspace{.5in}
{\bf Abstract} \bigskip \end{center} 
\setcounter{page}{0} 
We extend the Wilson renormalization group (RG) to supersymmetric
theories.  As this regularization scheme preserves supersymmetry, we
exploit the superspace technique. To set up the formalism we first
derive the RG flow for the massless Wess-Zumino model and deduce its
perturbative expansion.
We then consider N=1 supersymmetric Yang-Mills and show that the local
gauge symmetry --broken by the regularization-- can be recovered by
a suitable choice of the RG flow boundary conditions. We restrict our
analysis to the first loop, the generalization to higher loops
presenting no difficulty due to the iterative nature of the
procedure. Furthermore, adding matter fields, we reproduce the
one-loop supersymmetric chiral anomaly to the second order in the
vector field.
\end{titlepage}
%*********************************************************
%\newpage

\section{Introduction}

Perturbative calculations in field theories require a regularization
procedure to deal with ultraviolet divergences.  A powerful and simple
method is that of dimensional regularization which has the remarkable
feature of preserving gauge symmetry.  However, it is clear that this
regularization breaks supersymmetry, since fermionic and bosonic
degrees of freedom match only in fixed dimensions.  The only
modification of dimensional regularization compatible with
supersymmetry (the so-called dimensional reduction \cite{dred}) turns
out to be inconsistent \cite{incon}.  
The lack of a consistent
regularization scheme which manifestly preserves supersymmetry implies,
in particular, that superspace formalism can be used only with some
care since naive manipulations may lead to ambiguities
\cite{ambi}. Although dimensional reduction does not seem to cause any
practical difficulty and is extensively used \cite{jones} to perform
perturbative calculations, it is worthwhile to look for manifestly
supersymmetric regularizations which take advantage of the superspace
technique \cite{gates} and are free of ambiguities.

Recently a regularization procedure based on the Wilson
renormalization group (RG) has been formulated \cite{P,B} for scalar
field theories \cite{BDM,Mo} and gauge theories \cite{B,noi,Wett}.
With this method one introduces an ultraviolet (UV) and infrared (IR)
cutoff, $\L$ and $\L_0$ respectively, in the propagator so that all
Feynman diagrams become convergent in the UV region. The cutoff
effective action can be obtained through the standard procedure using
this modified propagator.  The physical effective action is recovered
by taking $\L=0$ and the limit \UV. The existence of such a limit is a
consequence of the renormalizability (and IR finiteness) of the
theory.  The RG formalism provides an alternative definition of the
cutoff effective action. Requiring the physical amplitudes to be
independent of $\L$, one derives a differential equation (the RG flow)
for the cutoff effective action, which can be solved giving a set of
boundary conditions that encode both renormalizability and the
renormalization conditions.  Despite the presence of the cutoff $\L$
which explicitly breaks gauge invariance, one proves that, by properly
fixing the boundary conditions of the RG equation, the Slavnov-Taylor
(ST) identity can be satisfied when all cutoffs are removed (at
least in perturbation theory).  This has been shown for the pure
Yang-Mills case both in terms of the ``bare'' couplings of the
effective action at the ultraviolet scale \cite{B} and of the physical
couplings
\cite{noi}.  Furthermore, as this method directly works in four
space-time dimensions, it naturally extends \cite{BV} to chiral gauge
theories with no ambiguity in the definition of the matrix $\gamma_5$
(contrary to dimensional regularization).

In this paper we generalize the RG method to supersymmetric
theories, implementing the regularization in such a way that
supersymmetry is preserved. This can be easily understood observing
how the cutoffs are introduced in the RG formulation.  One splits the
classical action into two parts, the quadratic and the interacting
one, and then multiplies the former by a cutoff function $\K(p)$ which
falls off sufficiently rapid for $p^2$ outside the region
$\L^2<p^2<\L_0^2$.  In the supersymmetric case it suffices to write
the classical action in terms of superfields and follow the same
procedure (in components this corresponds to use the same cutoff
function for all fields).  As our formulation works in $d=4$,
supersymmetry is maintained and, from the very beginning, we can
exploit the superspace technique, which simplifies perturbative
calculations and is now unambiguous.

For a supersymmetric gauge theory, supersymmetry is preserved though
gauge symmetry is explicitly broken by the regularization.  As for
non-supersymmetric gauge theories, we will show that by properly
fixing the boundary conditions of the RG flow the ST identity
associated to the gauge symmetry is recovered, when the matter
representation is anomaly free. However, if the matching conditions
for the anomaly cancellation are not fulfilled, we will be able to
reproduce the chiral anomaly.

The paper is organized as follows. In sect.~2 we set up the RG
formalism for the massless Wess-Zumino (WZ) model and, as an example
of how to perform perturbative calculations, we compute the one-loop
two-point function.  In this section we briefly discuss the boundary
conditions for the flow equation. In sect.~3 we consider the N=1 super
Yang-Mills (SYM) case in this framework.  In sect.~4 we formulate the
effective ST identity for this theory and show how to solve the fine-tuning
equation at one-loop order and at the UV scale.  Sect.~5 is devoted to
the computation of the one-loop chiral anomaly (up to the second order
in the gauge field) and sect.~6 contains some conclusions.  Finally,
our conventions are given in  the appendix.

\section{Renormalization group flow and effective action for the 
Wess-Zumino model}

The massless WZ model is described by the classical Lagrangian \cite{WZ}
$$
S_{cl}= S_2 + \si^{(0)}
$$
where 
\beeq &&
S_2=\frac{1}{16}\zint \bphi\; \phi\,, \qquad\qquad
\;\;\;\;\;\;\; \zint=\zzint\,,
\nonumber\\ &&
\si^{(0)}=
\frac1{48}\,\cint  \phi^3 + \mbox{h.c.}
\eeeq
and $\phi$ ($\bphi$) is a chiral (anti-chiral) superfield satisfying 
$\bD_\ad\phi=0$ ($D^\a\bphi=0$).

In order to quantize the theory one
needs a regularization procedure of the ultraviolet divergences.
We regularize these divergences by assuming that in the path integral
one integrates only the fields with frequencies smaller than a given
UV cutoff $\L_0$. This procedure is equivalent to modify the free
action to assume that the free
propagators vanish for $p^2 > \L_0^2$.

The generating functional is
\beq\nome{W}
Z[J]=e^{iW[J]}=\int {\cal D}\Phi  \, \exp{i\lgr\half (\Phi, \,
\DD^{-1} \Phi)_{0\L_0}+(J,\Phi)_{0\L_0}+\si[\Phi;\L_0]\rgr}
\,,
\eeq
where we have collected the fields and the sources in $\Phi_i=(\phi,\,\bphi)$
and $J_i=(J,\,\bJ)$ respectively,
and we introduce the general  cutoff scalar products between 
fields and sources
\beeq&&
\half (\Phi,\,\DD^{-1}\Phi)_{\L\L_0}\equiv\frac{1}{16}
\spint\, K^{-1}_{\L\L_0}(p)\,
\bphi(-p,\,\th)\,\phi(p,\,\th)\,,
\,\;\;\;\;
\spint \equiv \int \frac{d^4p}{(2\pi)^4} \,d^2\th \,
d^2\bt
\nome{prop}
\\&&
\nonumber
\\&&
(J,\Phi)_{\L\L_0}\equiv\frac{1}{16}
\spint\, K^{-1}_{\L\L_0}(p) \,
\lgr J(-p,\,\th)\,\frac{D^2}{p^2}\,\phi(p,\,\th)\,
+\bJ(-p,\,\th)\,\frac{\bD^2}{p^2}\,\bphi(p,\,\th)
\rgr
\,,
\eeeq
with $K_{\L\L_0}(p)$ a cutoff function which is one for $\L^2 < p^2 <
\L_0^2$ and rapidly vanishes outside~\footnote{The factors
$D^2/(16p^2)$, $\bD^2/(16p^2)$ are needed to write the chiral and
anti-chiral superspace integral respectively, as integrals over the
full superspace (see the appendix).}.  The introduction of such a cutoff
function in \re{prop} yields a regularized propagator which preserves
supersymmetry, this being a global, linearly realized
transformation. Hence the UV action $\si[\Phi;\L_0]$ in \re{W}
contains all possible renormalizable supersymmetric interactions, \ie
superspace integrals of superfields and their covariant derivatives
which are local in $\th$.  Dimensional analysis tells us that they are
given by the monomials $\phi\bphi$, $\phi\,,\phi^2\,, \phi^3$,
$\bphi\,,\bphi^2\,, \bphi^3$, properly integrated.

According to Wilson one integrates over the fields with frequencies
$\L^2<p^2<\L_0^2$ and obtains \cite{P,B}
\beq\nome{Z'}
e^{iW[J]}=N[J;\L]\int {\cal D}\Phi \, 
\exp{i\lgr \half(\Phi, \,\DD^{-1} \Phi)_{0\L}+(J,\Phi)_{0\L}
+\se[\Phi;\L]\rgr}
\,,
\eeq
where $K_{0\L}=K_{0\L_0}-\K$ and $N[J;\L]$ contributes to the
quadratic part of $W[J]$.  The functional $\se[\Phi;\L]$ is the 
so-called Wilsonian effective action\footnote{Here and in the following
we explicitly write only the dependence on the cutoff $\L$, since the
theory is renormalizable and we are interested in the limit \UV.}.
Since the regularization preserves supersymmetry, this functional
contains all possible supersymmetric interactions.  One can show that
$\se[\Phi;\L]$ is the generating functional of the connected amputated
cutoff Green functions --- except the tree-level two-point function
--- in which the free propagators contain $\L$ as an infrared cutoff
\cite{BDM}.  In other words the functional
\beq\nome{WL}
W[J';\L]= 
\se[\Phi;\L]+\half (\Phi,\, \DD^{-1}\Phi)_{\L\L_0} \,,
\eeq
with the sources $J'$ given by 
\beq\nome{J'}
J_i'(-p,\th)=\K^{-1}(p)\,D^{2\ej}(\th)\Phi_j(-p,\th)\, \DD^{-1}_{ji}(p)\,,
\eeq
is the generator of the cutoff connected Green functions.  The matrix
$\DD^{-1}_{ij}$ is defined through \re{prop} and its entries are
$\DD^{-1}_{ij}=1/16$ if $i \ne j$ and zero otherwise. Moreover, in
order to keep formulas more compact, we have introduced the
two-component vector $\eps_k = \lp 1, -1\rp$ and the  shortened notation
$D^{-2}\equiv\bD^2$ which allow to treat simultaneously chiral and
anti-chiral fields.

\subsection{Evolution equation}
The requirement that the generating functional \re{Z'} is independent 
of the IR cutoff $\L$ gives rise to a differential equation for the
Wilsonian effective action, the so-called exact RG equation
\cite{P,B}, which can be translated into an equation for $W[J;\L]$ 
\beq\nome{eveqW}
\L\p_{\L} W[J;\L]=\half\spint \L\p_{\L} \K^{-1}(p)\,\DD^{-1}_{ij}(p)
\lp
\frac{\de W}{\de J_i(-p,\th)}\frac{\de W}{\de J_j(p,\th)}
-i \,\frac{\de^2 W}{\de J_i(-p,\th) \de J_j(p,\th)}
\rp
\,.
\eeq
This equation can be more easily understood   
taking into account that  $\L$ enters as an IR cutoff in
the internal propagators of the cutoff Green functions. 

It is convenient to introduce the so-called ``cutoff effective
action'' which is given by the Legendre transform of $W[J;\L]$
\beq\nome{Leg}
\G[\Phi;\L]=W[J;\L]-\cint J\phi -\acint \bJ\bphi\,.
\eeq
This functional generates the cutoff vertex functions in which the 
internal propagators have frequencies in the range $\L^2<p^2<\L_0^2$ 
and reduces to
the physical quantum effective action in the limits $\L\to 0$ and \UV
\cite{BDM,Mo}.

The evolution equation for the functional $\G[\Phi;\L]$ can be derived
from \re{eveqW} by using \re{Leg} and inverting the functional
$\frac{\de^2 W}{\de J\de J}$.  This inversion can be performed
isolating the full two-point contributions $\G_2$ in the
functional $\G[\Phi;\L]$
$$
(2\pi)^8 \frac{\de^2\G}{\de\Phi_j(p,\th_1)\,\de\Phi_k(k,\th)}
= (2\pi)^4 \G_{2\;kj}(k;\L) \, D^{-2\ek}(\th)\,  
D^{-2\ej}(\th_1)\,
\sde (k+p) + \Gi_{kj}[\Phi;k,p;\L] 
$$
and  $W_2$ in $W[J;\L]$
\beq\nome{inversionw}
(2\pi)^8 \frac{\de^2 W}{\de J_k(-k,\th)\,\de J_{i}(q,\th_2)}
= (2\pi)^4 W_{2\;ik}(k;\L) \, D^{-2\ei}(\th_2)\,  
D^{-2\ek}(\th)\,
\sde (q-k) + \Wi_{ik}[J;q,-k;\L]
\eeq
where the dependence on Grassmann variables in $\Gi$ and $\Wi$ is
understood. Henceforth we will prefer writing all integrals in the
full superspace, so that we have to cope with factors like
$\frac{D^2(\th)}{16 k^2}$ and $\frac{\bD^2(\th)}{16 k^2}$ originating
from chiral and anti-chiral projectors, respectively.  
These two factors can be simultaneously treated  with the help of the
vector $\ek$ and identifying $\lp\frac{D^2(\th)}{16k^2}\rp^{-1}$ with
$\frac{\bD^2(\th)}{16 k^2}$.

Then making use of the identity 
\beeq
\dfud{\Phi_i(-q,\,\th_2)}{\Phi_j(p,\,\th_1)} &=& D^{-2\ei}(\th_1)\, 
\sde (q+p) \, \de_{ij}
\nonumber \\
&=& (2\pi)^8 \skint 
\frac{\de^2 W}{\de J_k(-k,\th) \de J_{i}(q,\th_2)}
\lp \frac{D^2(\th)}{16 k^2}\rp^{\ek} 
\frac{\de^2\G}{\de\Phi_j(p,\th_1) \de\Phi_k(k,\th)}\no
\eeeq
we can express $\Wi_{ij}$ in \re{inversionw} as a functional of 
$\Phi$ obtaining
\beq \nome{wint}
\Wi_{ij}[J(\Phi);q,\,p;\L]= -\G_{2\;lj}^{-1}(p;\L)\lp 
\frac{D^2(\th_2)}{16\,q^2}\rp^{\ek}
\,\lp\frac{D^2(\th_1)}{16\,p^2}\rp^{\el}\,\bG_{kl}[\Phi;q,\,p;\L]\,
\G_{2\;ik}^{-1}(q;\L)
\,,
\eeq
where the auxiliary functional $\bG$ satisfies the recursive equation
\beq \nome{gammab}
\bG_{ij}[\Phi;q,p;\L]= (-)^{\de_j}\Gi_{ij}[\Phi;q,p;\L]-\skint 
\lp\frac{1}{16k^2}\rp^{\abs{\ek}}
\,\Gi_{kj}[\Phi;k,p;\L]\,\G_{2\;lk}^{-1}(k;\L)\,\bG_{il}[\Phi;q,-k;\L]
\eeq
which gives $\bG$ in terms of the proper vertices of $\G$.  The
grassmannian parity $\de_j$ is zero for the (anti)chiral superfield
and the factor $(-)^{\de_j}$  has been introduced to take into 
account the possible anti-commuting nature of the field (it will be 
needed in SYM).

Finally, inserting \re{inversionw}  in \re{eveqW} and using \re{wint},
we obtain the evolution equation for the functional $\G_{\L} [\Phi]$ 
\beeq \nome{eveq}
&&\L\p_{\L}\lq \G_{\L} [\Phi]-
\half \,\spint \K^{-1}(p)\, \Phi_i(-p,\th)
\,\DD^{-1}_{ij}(p)\,\Phi_j(p,\th)
\rq
=-\frac i 2 \sqint \L\p_{\L} \K^{-1}(q)
\nonumber \\ 
&&\;\;\;\;\;\;\;\;\;\;\;\;\;\times\,
\G_{2\;lj}^{-1}(q;\L)\;
\DD^{-1}_{ji}\;
\G_{2\;ik}^{-1}(q\;\L)
\lp\frac{D^2(\th)}{16\,q^2}\rp^{\ek}
\lp\frac{D^2(\th)}{16\,q^2}\rp^{\el}
\bG_{kl}[\Phi;q,\,-q;\L]\,.
\eeeq
This equation, together with a set of suitable boundary conditions,
can be thought as an alternative definition of the theory which in
principle is non-perturbative. As far as one is concerned with its
perturbative solution, the usual loop expansion is recovered by
solving iteratively \re{eveq}. Such a solution is possible since the
l.h.s. of \re{eveq} at a given loop order depends only on lower loop
vertices. The RG formulation provides
a very simple method to prove perturbative renormalizability, \ie that the
\UV limit can be taken. The proof is a straightforward generalization of 
that given in \cite{P}-\cite{BDM}  for non-supersymmetric theories.

\subsection{Relevant couplings and boundary conditions}
In order to set the boundary conditions one distinguishes between 
relevant  couplings and irrelevant vertices according to their mass dimension.
Relevant couplings have non-negative mass dimension and are 
defined as the value of some vertices and their derivatives 
at a given normalization point. 
Dimensional analysis tells us that they originate from the monomials 
$\phi\bphi$, $\phi\,,\phi^2\,, \phi^3$, $\bphi\,,\bphi^2\,, \bphi^3$,
properly integrated. 

The massless chiral multiplet two-point function 
(\ie the $\phi\bphi$-coefficient of the cutoff effective action)
\beq \nome{gamma2}
\G_{2\;ij}(p;\L)=\DD^{-1}_{ij}\K^{-1}(p)+\S_{2\;ij}(p;\L)
\eeq
contains the relevant coupling 
$$
Z_{ij}(\L)=\S_{2\;ij}(p;\L) 
\left. \right|_{p^2=\mu^2}$$
where $\mu$ is some non-vanishing subtraction point, whose
introduction, being $\phi$ a massless field, is required to avoid 
the IR divergences.
We need not define the remaining relevant couplings since the
corresponding monomials are not generated in perturbation theory
(as can be seen from \re{eveq}, only vertices with an equal number
of $\phi$ and $\bphi$ receive perturbative contributions).

All the vertices appearing  
with a number of $\phi\bphi$ larger than one are irrelevant.   
A further contribution to the irrelevant part of $\G$ comes from 
the two-point function,  and is given by 
$$
\S^{\mbox{\scriptsize{irr}}}_{2\;ij}(p;\L)\equiv\S_{2\;ij}(p;\L)-Z_{ij}(\L)\,.
$$

One assumes the following boundary conditions: 

\noindent (i) at the UV scale $\L=\L_0$ all irrelevant vertices
vanish. As a matter of fact $\G[\Phi;\L=\L_0]$ reduces to the bare
action, which must contain only renormalizable interactions  
in order to guarantee perturbative renormalizability;

\noindent (ii)  the relevant couplings are fixed at the physical
point $\L=0$ in terms of the physical couplings, such as the wave 
function normalization, the three-point coupling and the mass. 
In the case at hand only the first coupling evolves with $\L$ 
whereas the remaining ones coincide with their tree-level value 
(\eg $m=0$ and $\l$) and for this reason are not investigated.

Hence the boundary conditions to be imposed  on the relevant and 
irrelevant part of $\G_{2\;ij}$ are
\beq\nome{bc}
Z_{ij}(\L=0)=0 \,,\;\;\;\;\;\;\;\;
\S^{\mbox{\scriptsize{irr}}}_{2\;ij}(p;\L_0)=0\,,
\eeq
respectively. 

\subsection{Loop expansion}

\noindent{\it (i) Tree level}

\noindent
The starting point of the iteration is the tree-level 
interaction
\beq\nome{gammaint}
\Gio_{ij}[\Phi;q,\,p;\L]=\frac{\l}{8}\, \de_{ij} \,\sppint
\de^4(\th_1-\th') D(\th')^{-2\ei}
\de^4(\th_2-\th')\,\Phi_j(p')\,\de^4(p+q+p')
\eeq
together with the tree-level two-point function 
$\G_{2\;ij}^{(0)}(p;\L)=\DD^{-1}_{ij} \K^{-1}(p)$. Inserting these
expressions in \re{gammab} one obtains the tree-level functional 
$\bG_{ij}^{(0)}[\Phi]$.

\noindent{\it (i) One-loop calculations}

\noindent
The evolution equation for the functional $\G[\Phi]$ at one-loop order can
be derived by writing the r.h.s of \re{eveq}  in terms of the known objects
$\bG_{ij}^{(0)}[\Phi]$ and $\G_{2\;ij}^{(0)}$.
For instance, the evolution equation for the two-point function 
is determined by the $\phi\bphi$-coefficient in \re{eveq} which, at the
tree level, originates only from the second term in the r.h.s. of 
\re{gammab}, \ie 
$$
-\skint 
\,\Gio_{ml}[\Phi;k,\,q;\L]\,\frac{\K(k)}{16k^2}\DD_{nm}(k)\,
\Gio_{kn}[\Phi;-q,\,-k;\L]
\,.
$$ Next, substituting \re{gammaint} in the expression above and
carrying out some standard $D$-algebra manipulations (reported in the
appendix), one finds
\beeq\nome{gamma21}
&& \spint \bphi (-p,\,\th)\, \L \p_\L \S_2^{(1)} (p;\, \L)
 \, \phi  (p,\,\th)\,=
\frac i{64}\l^2\,
\spqint
\frac {\K(p+q)\L \p_\L\K(q)}{q^2(p+q)^2}\, \nonumber \\
&&\;\;\;\;\;\;\;\;\;\;\;\;\;\times\,
\bphi (-p,\,\th_1)\, \phi  (p,\,\th_2)\, \de^{4} (\th_1-\th_2)\,
\bD^2 D^2(q,\,\th_2)  \, \de^{4} (\th_1-\th_2)\,.
\eeeq
Notice that eq.~\re{eveq} describes only the evolution of the interacting
part of $\G$, since the tree level in \re{gamma2} cancels out.

Recalling the property
\beq\nome{susydelta}
 \de^{4} (\th_1-\th_2)\,\bD^2\,D^2  \, \de^{(4)} (\th_1-\th_2)=
 \de^{4} (\th_1-\th_2)\,,
\eeq
one gets
\beq\nome{sigma2}
\L \p_\L  \S_2^{(1)} (p;\, \L) =\frac i{128} \,\l^2\,
\int \frac{d^4 q}{(2\pi)^4}
 \; \frac {\L \p_\L(\K(q)\,\K(p+q))}{q^2\,(p+q)^2}\,.
\eeq
Implementing  the  boundary conditions \re{bc}, the solution of 
\re{sigma2} at the physical point $\L=0$ and in the \UV limit is 
$$
\S_2^{(1)} (p;\, \L=0) =\frac i{128} \,\l^2\,
\int \frac{d^4 q}{(2\pi)^4}
 \; \lp\frac {1}{q^2\,(p+q)^2}-\left.\frac
{1}{q^2\,(p+q)^2}\right|_{p^2=\mu^2}\rp
\,.
$$
Notice the crucial role of the boundary condition for $Z_{ij}$, \ie 
$Z^{(1)}_{ij}(0)=0$,
which naturally provides the necessary 
subtraction to make the vertex function $\S_{2\;ij}$ finite for \UV. 
Conversely  one can see from power counting that the remaining 
irrelevant vertices are finite, and no subtraction is needed.
This property holds at any order in perturbation theory \cite{BDM}.

\section{N=1 Super Yang-Mills}

The super Yang-Mills (SYM) action reads~\cite{sym-orig} (the
conventions are those of~\cite{piglect})
$$
\sym  = -\frac{1}{128g^2}\tr\cint \WW^\a \WW_\a\,, \;\;\;\;\;\;\;\;
\WW_\a = \bD^2\lp e^{-gV}D_\a e^{gV}\rp\,,
$$
where $V(x,\th)$ is the $N=1$ vector supermultiplet which belongs to
the adjoint representation of the gauge group $G$. In the matrix
notation $V=V^a\tau_a$, 
with  the matrices $\tau_a$ satisfying 
$[\tau_a,\tau_b]=if_{abc}\tau_c\,$,$\tr\tau_a\tau_b=\de_{ab}$.
The classical action is invariant under the gauge transformation
\beq
e^{gV'} = e^{-i\bchi}e^{gV} e^{i\chi}\, ,\quad\quad\quad  \bD_\ad\chi=0\,,
\quad  D^\a\bchi=0\,,
\eeq
where $\chi=\chi^a\tau_a$.

In order to quantize the theory one has to fix the gauge and choose a
regularization procedure. From what we have seen so far it should be
manifest that the introduction of the cutoff
does not spoil global symmetries as long as they are linearly
realized. If this is not the case the transformation of the quadratic
part of the action mixes with the transformation of the rest (recall that the
cutoff function multiplies only the quadratic part of the classical action).
Therefore, we shall choose a supersymmetric gauge fixing
instead of the familiar Wess-Zumino one in which the
supersymmetry transformation is not linear.

As described in ref.~\cite{gates}, we add  to the action a 
gauge fixing term which is a supersymmetric extension of the
Lorentz gauge and the corresponding Faddeev-Popov term
\beeq
&&\sgf=-\frac{1}{128\a}\tr\zint D^2V\bD^2V
\nonumber\\
&&\sfp= -\frac{1}{8}\tr\zint\lp\cm +\bcm\rp \lq
\half L_{gV}(\cp+\bcp)+ \half 
\lp L_{gV}{\rm coth}(L_{gV}/2)\rp\lp\cp-\bcp\rp \rq
\nonumber
\\&&\phantom{\sfp}=
 -\frac{1}{8}\tr\zint\lp\cm +\bcm\rp\lq 
\cp -\bcp +\half g\,[V,\cp+\bcp]+\cdots \rq\,,
\eeeq 
where  the ghost $\cp$ and the anti-ghost $\cm$ are chiral fields,
like the gauge  parameter $\chi$, and $L_{gV}\,\cdot=[gV,\cdot]$.
The classical action
$$
\scl=\sym+\sgf+\sfp
$$
is invariant under the BRS transformation 
\beeq
&&
\de V = \eta \lq
\half L_{gV}(\cp+\bcp)+ \half 
\lp L_{gV}{\rm coth}(L_{gV}/2)\rp\lp\cp-\bcp\rp \rq \,,
\nonumber \\[1mm]&&
\de\cp= -\eta\, \cp^2\,, \;\;\;\;\;\;\;\;\;\;\;\;\;\;\;\;\;\;\;\;
\de\bcp=-\eta \,\bcp^2\,,
\nonumber \\&&
\de\cm= - \eta \frac1{16\a}\bD^2 D^2V \,, \;\;\;\;\;
\de\bcm= - \eta \frac1{16\a} D^2\bD^2 V\no
\eeeq
with $\eta$ a Grassmann parameter. Introducing the sources 
$\g_i=(\gv$, $\g_{\cp}$,
$\g_{\bcp})$, 
associated to the BRS variations of
the respective superfields, the BRS action in the Fermi-Feynman gauge
($\a=1$) reads 
\beeq \nome{brsac}
&&\sbrs =\scl +\zint 
\gv \lq \half L_{gV}(\cp+\bcp)+ \half 
\lp L_{gV}{\rm coth}(L_{gV}/2)\rp\lp\cp-\bcp\rp \rq  
\nonumber \\&&
\phantom{\sbrs =\scl}
-\cint \g_{\cp} \cp^2 - \acint \g_{\bcp}\bcp^2 
\nonumber \\&&
\phantom{\sbrs}
= S_2 +\si^{(0)}
\eeeq
with 
$$
S_2=\zint\lq\frac{1}{16}V\partial^2V +\frac{1}{8}\lp\cm\bcp -\bcm\cp\rp\rq\,.
$$
Notice that in \re{brsac} we did not introduce the BRS 
sources for $\cm$ and $\bcm$
since one can show that the effective action depends on these fields
and the source $\gv$ only through the combination
$$
\tgv=\gv-\frac18\lp \cm+\bcm\rp\,.
$$

As described in the previous section for the WZ model, we
regularize the UV divergences multiplying the free propagators 
by a cutoff function $\K$, so that the generating functional 
$Z[J,\g]$ can be written as in \re{W}
with 
$$
\Phi_i= (V,\, \cp,\,         \bcm,    \,  \cm,   \, \bcp)\,,\;\;\;\;\;
J_i=(J_V, \, \xi_{-}+\bD^2\gv,\, -\bxi_{+},\, -\xi_{+},\, \bxi_{-}-
D^2\gv)
$$
and the cutoff scalar product between fields and sources given by
\beeq\nome{phiphi}
&&
(\Phi,\,\DD^{-1}\Phi)_{\L\L_0}=\spint \K^{-1}(p)
\lgr-\frac{1}{16}V(-p,\th)\,p^2V(p,\th) \right.
\nonumber\\&&
\phantom{(\Phi,\,\DD^{-1}\Phi)_{\L\L_0}=\spint \K^{-1}(p)}\left.
+\frac{1}{8}\lq\cm(-p,\th)\bcp(p,\th)
-\bcm(-p,\th)\cp(p,\th)\rq\rgr
\eeeq
and
\beeq\nome{jphi}
&&(J,\Phi)_{\L\L_0}=\spint \K^{-1}(p)\lgr J_V(-p,\th)\,V(p,\th)+
\frac1{16}\lq
\lp\xi_{-}+\bD^2\gv\rp(-p,\th)\,\frac{D^2}{p^2}\cp(p,\th) 
\nonumber \right. \right. \\ && 
\phantom{(J,\Phi)_{\L\L_0}=\spint \K^{-1}(p)}
+\frac{\bD^2}{p^2}\bcm(-p,\th)\,\bxi_{+}(p,\th)+\frac{D^2}{p^2}\cm(-p,\th)
\,\xi_{+}(p,\th)
\nonumber \\ && 
\phantom{(J,\Phi)_{\L\L_0}=\spint \K^{-1}(p)}
+\left. \left.
\lp\bxi_{-}-D^2\gv\rp(-p,\th)\,\frac{\bD^2}{p^2}\bcp(p,\th)\rq \rgr\,. 
\eeeq
The UV action 
$\si[\Phi,\g;\L_0]$ contains all possible relevant interactions 
written in  terms of $\Phi_i$, $\g_i$ and superspace derivatives,
which are invariant under Lorentz and global gauge transformations. 
Notice that at the tree level all quadratic contributions in the
fields and sources are gathered in \re{phiphi} and \re{jphi}. 

Afterwards one integrates over the fields with frequencies
$\L^2<p^2<\L_0^2$ and the result is the  analogue of \re{Z'}
where the Wilsonian effective action $\se[\Phi,\g;\L]$ depends also on
the BRS sources.
The generating functional of the cutoff connected Green functions
$W[J,\g;\L]$ is given by \re{WL} and \re{J'} with 
$\ek$ the five-component vector $\ek=(0,1,-1,1,-1)$ and 
the matrix $\DD^{-1}_{ij}$ defined through \re{phiphi}.  
This matrix turns out to be block-diagonal and its entries are
$1/8(-p^2,\, \eps_{\mbox{\tiny{AB}}},\,  
\eps_{\mbox{\tiny{AB}}})$, ${\mbox{\footnotesize{A}}}=(+,-)$, 
with  $\eps_{\mbox{\tiny{AB}}}=-\eps_{\mbox{\tiny{BA}}}$ and 
$\eps_{+-}=1$.
The derivation of the evolution equation for the functional $W$
exactly follows that of the WZ model presented in sect.~2. 
Finally the cutoff effective action $\G$ 
\beeq \nome{Legsym}
&&\G[\Phi,\g;\L]=W[J,\g;\L]-\zint J_V V -\cint \lp \xi_-\cp + \cm
\xi_+ \rp  
\nonumber \\ &&
\phantom{\G[\Phi,\g;\L]=W[J,\g;\L]}
-\acint  \lp \bxi_-\bcp + \bcm \bxi_+ \rp 
\eeeq
evolves according to \re{eveq} with the appropriate vertices, $\DD_{ij}$ 
and $\ek$. 

\subsection{Matter fields}
When adding matter fields to the pure  super Yang-Mills action one
gets SQCD, the supersymmetric generalization of QCD.
Matter is described by a set of chiral superfields 
$\phi^{\mbox{\tiny{I}}}(x,\th)$  which
belong to some representation $R$ of the gauge group. 
Their BRS transformation reads
$$
\de \phi^{\mbox{\tiny{I}}}= -\eta\,\cp^a \,
T_a{}^{\mbox{\tiny{I}}}{}_{\mbox{\tiny{J}}} \,\phi^{\mbox{\tiny{J}}} \equiv
-\eta(\cp\phi)^{\mbox{\tiny{I}}}  
\,,\quad 
\de \bphi_{\mbox{\tiny{I}}} =  \eta\,\bphi_{\mbox{\tiny{J}}} 
\,T_a{}^{\mbox{\tiny{J}}}{}_{\mbox{\tiny{I}}}\,\bcp^a 
\equiv \eta(\bphi\, \bcp)_{\mbox{\tiny{I}}} \, ,
$$
where the hermitian matrices $T_a$ are the generators of the gauge
group in the representation $R$. 

The BRS action for the matter fields is
\beq \nome{matter-action}
S_{\rm matter} = \frac{1}{16} \zint  \bphi\, e^{gV^a T_a}\phi - 
\cint \g_\phi\,  \cp\,\phi + \acint \g_{\bphi}\, \phi\,\bcp  
\eeq
plus a possible superpotential $W$ having the general form 
$W(\phi) = \frac{1}{8}m_{({\mbox{\tiny{IJ}}})}\phi^{\mbox{\tiny{I}}} 
\phi^{\mbox{\tiny{J}}} + \l_{({\mbox{\tiny{IJK}}})}\phi^{\mbox{\tiny{I}}} 
\phi^{\mbox{\tiny{J}}}
\phi^{\mbox{\tiny{K}}}$,
the mass matrix $m_{{\mbox{\tiny{IJ}}}}$ and the Yukawa coupling constants 
$\l_{{\mbox{\tiny{IJK}}}}$
being invariant symmetric tensors in the representation $R$.

Developing the RG formalism in presence of matter
fields is straightforward once one replaces the sets of fields and sources
with 
\beeq 
&&
\Psi_i= (V,\, \cp,\,         \bcm,    \,  \cm,   \, \bcp\,\phi,\,\bphi)
\,,\;\;\;\;\; \g_i=(\gv\,,\g_{\cp}\,,\g_{\bcp}\,,\g_\phi\,,\g_{\bphi})\,,
\nonumber \\ 
&&
J_i=    (J_V, \, \xi_{-}+\bD^2\gv,\, -\bxi_{+},\, -\xi_{+},\, \bxi_{-}-
D^2\gv,\,J,\,\bJ)\,.
\eeeq
The evolution equation for the effective action 
has the usual form \re{eveq}, with a natural redefinition of $\eps_k$
and $\DD^{-1}_{ij}$  to take into account matter fields
(\eg $\,\eps_k = \lp 0,\,1,\, 1,\, -1,\, -1,\, 1, -1 \rp$).

\subsection{Boundary conditions} 
As discussed in subsect.~2.2 we first distinguish between relevant and 
irrelevant vertices.  The relevant part of the
cutoff effective action involves full superspace integrals 
of monomials in the fields, sources and derivatives local in $\th$ and
with dimension not larger than two 
\beq\nome{gammarel}
\Gr[\Psi,\g;\s_i(\L)]=\sum_i \s_i(\L)\,P_i[\Psi,\g]\,,
\eeq
where the sum is over the monomials $P_i[\Psi,\g]$ invariant under
Lorentz and global gauge transformations. Due to the dimensionless nature of
the field $V$ this sum contains infinite terms which can be classified
according to the number of gauge fields.
The couplings $\s_i(\L)$ can be expressed in terms of the 
cutoff vertices at a given subtraction point, generalizing the
procedure used in  subsect.~2.2 to define the coupling $Z_{ij}(\L)$    
(see  also \cite{noi} for the technique  of
extracting the relevant part from a given functional with a
non-vanishing subtraction point in the non-supersymmetric Yang-Mills case).

The remaining part of the cutoff effective action 
is called ``irrelevant''. 
The  boundary condition we impose on  the 
irrelevant part of the cutoff effective action is
$$
\Gir[\Phi,\g;\L=\L_0]=0\,.
$$
For $\L=\L_0$, then, the cutoff effective action becomes ``local'',
\ie an infinite sum of local terms,  and
corresponds to the UV action $\si[\Psi,\g;\L_0]$, with the bare
couplings given by  $\s_i(\L_0)$. 

The way in which the boundary conditions for the relevant couplings
$\s_i(\L)$ are determined is not straightforward.  In sect.~2 we fixed
them at the physical point $\L=0$ in terms of the value of the
physical couplings (such us the normalization of the chiral field).
In the case of a gauge theory, as the one we are considering, there
are interactions in \re{gammarel} which are not present in $\sbrs$, so
that only some of the relevant couplings are connected to the physical
couplings (such as the wave function normalizations and the
three-vector coupling $g$ at a subtraction point $\mu$).  For instance
the contribution to \re{gammarel} with two gauge fields consists of
three independent monomials 
$$
\zint \tr \lq \s_1\, V\,V 
+ \s_2\,V\,D^\a\,\bD^2\,D_\a\, V 
+ \s_3\,V\,D^2\,\bD^2\,V\rq
$$
instead of the two in $\sbrs$.
Therefore, in order to fix the boundary conditions for all the
relevant couplings, one needs an additional  fine-tuning procedure which
implements  the gauge symmetry at the physical point.
However, this analysis involves non-local functionals and 
is highly not trivial.
Alternatively one can discuss the symmetry at the ultraviolet scale
and determine $\s_i(\L=\L_0)$.  In this case the discussion is
simpler, since all functionals are relevant, but one has to perform a
perturbative calculation (\ie to solve the RG equations)  to obtain
the physical couplings. Notice that while the physical couplings 
are independent of the cutoff function, the bare action, \ie
$\s_i(\L_0)$, is generally not. 

In this paper we consider the second possibility,
although  the wave function normalizations and the gauge coupling $g$
at a subtraction point $\mu$ are still set at $\L=0$. As a matter of
fact  there are combinations of the monomials in \re{gammarel} which 
are not involved in the fine-tuning, so that the corresponding 
couplings are free and can be fixed at the physical point $\L=0$.
Before explaining the details of the fine-tuning procedure we recall how 
to implement the gauge symmetry in the RG formulation.

\section{Effective ST identity}

The gauge symmetry requires that the physical effective action
satisfies the ST identity \cite{BRS}
\beq\nome{ST0}
\SS_{\G'}\G'[\Psi,\g]=0\,,
\eeq
where
$\G'[\Psi,\g]=\G[\Psi,\g]+\frac1{128} \tr \zint D^2 V\bD^2 V$
and~\footnote{From now on the sum over the fields in $\Psi$ will not include
$\cm$ and $\bcm$.}
\beq \nome{ST}
\SS_{\G'}=\spint
\lq\lp \frac{D^2}{16p^2} \rp^{\ei}
 \dfud{\G'}{\Psi_i(-p)}\,
\dfud{}{\g_i(p)} +
\lp \frac{D^2}{16p^2} \rp^{\ei} 
\dfud{\G'}{\g_i(p)}\,
\dfud{}{\Psi_i(-p)}\rq
\eeq
is the Slavnov operator. The ST identity can be
directly formulated for the Wilson effective action $\se$ at any $\L$.
Consider the generalized BRS transformation
\beq\nome{brseff}
\de\Psi_i(p)=\Ki(p)\,\eta\,\dfud{\st}{\g_i(-p)}\,, \;\;\;\;\;
\de\cm= - \eta \frac1{16}\bD^2 D^2V \,, \;\;\;\;\;
\de\bcm= - \eta \frac1{16} D^2\bD^2 V\,,
\eeq
where $\eta$ is a Grassmann parameter and $\st$ is the total action
(\ie $\se$ plus the source and the quadratic terms in \re{Z'}). 
Performing such a change of variable in the
functional integral \re{Z'}, one deduces the following identity
\beq \nome{STeff}
\SS_J Z[J,\g]=N[J,\g;\L] \int \DD\Psi \exp{i\lgr
\half(\Psi,\DD^{-1} \Psi)_{0\L}+(J,\Psi)_{0\L} 
+\se[\Psi;\L]\rgr} \, \De[\Psi,\g;\L]\, ,
\eeq
where $\SS_J$  is the usual ST operator
$$
\SS_J=\spint J_i(p)\,(-)^{\de_i}\dfud{}{\g_i(p)}+
\frac1{16} \spint \lq D^2 \xi_+ (p)+\bD^2 \bxi_+ (p)\rq
\dfud{}{J_V(p)}
$$
with ${\de_i}$  the source ghost number, and the functional $\De$ reads:
$$
\De[\Psi,\g;\L]=
i \spint  \Ki(p)\,\exp{(-i\se)}\lgr \dfud{}{\Psi_i(p)}
\,\dfud{}{\g_i(-p)}\rgr \exp{(i\se)}\qquad\qquad\qquad\;\;
$$
$$
-i\spint  \lq \Psi_i(p)\,\DD^{-1}_{ij}(p)\,
\dfud{}{\g_j(p)}+(\cp-\bcp)(p)\dfud{}{V(p)}
-\frac1{16} V(p)\Bigl(D^2 \dfud{}{\cm(p)}+\bD^2  \dfud{}{\bcm(p)}\Bigr)
\rq \se
\,.
$$
Whereas the l.h.s of the identity \re{STeff} arises from the variation of
the source term $(J,\Psi)_{0\L}$, the functional $\De$ originates from
the Jacobian of the transformation \re{brseff} and from
the variation of the rest of $\st$.
Restoration of symmetry, $\SS_J Z[J,\g]=0$, translates into 
$$
\De[\Psi,\g;\L]=0 \;\;\;\; {\mbox{for any}} \;\;\L\,.
$$
From a perturbative point of view, instead of studying $\De$ 
it is convenient to introduce 
\cite{BV,MT,Ellw} its Legendre transform $\DG$, in which reducible
contributions are absent. 
Recalling \re{WL} and \re{J'} which relate $\se[\Psi,\g;\L]$ 
to $W[J,\g;\L]$, and using \re{Leg}, \re{Legsym} one finds 
$$
\DG[\Psi,\g;\L]=-\spint\lq \Kiu(p) \lp \frac{D^2(\th_1)}{16\,p^2}
\rp^{\ei} 
\dfud{\G'}{\Psi_i(-p)}\, \dfud{\G'}{\g_i(p)}
-\frac{\Ki(p)}{\K(p)}\,\DD^{-1}_{ij}(p)\,
 \Psi_i(p) \dfud{\G'}{\g_i(p)}\rq
$$
\beeq
&&
-i\,\hbar \spqint \frac{\Ki(p)}{\K(p)}\,\DD^{-1}_{ij}(p)
\lp \frac{D^2(\th_2)}{16 q^2}
\rp^{\ek} (-)^{\de_i}
\frac{\de^2 W}{\de J_i(p) \de J_k(q)} \qquad\qquad
\no\\&&
\phantom{-i\hbar \spqint \frac{\Ki(p)}{\K(p)} }
\times\frac{\de^2}{\de\Psi_k(-q) \de \g_j(-p)} 
\lp \G -\zint \gv (\cp-\bcp) \rp \,,
\eeeq 
where $\de^2W/\de J\de J$ is that functional of $\Psi$ and $\g$
appearing in the inversion \re{inversionw} and \re{wint}.
Finally, after performing such an inversion, 
the cutoff ST identity reads
\beq\nome{DG}
\DG[\Psi,\g;\L]\equiv\DGb +\DGh=0\,,
\eeq 
with
\beq \nome{dgb'}
\DGb=
-\spint \Kiu(p) 
\lp \frac{D^2(\th_1)}{16p^2} \rp^{\ei} 
\dfud{\G'}{\Psi_i(-p)}\dfud{\G'}{\g_i(p)}
+\spint \frac{\Ki(p)}{\K(p)}\DD^{-1}_{ij}(p)
\Psi_i(p) 
\dfud{\G'}{\g_i(p)}
\eeq
and
\beeq\nome{dgh}
&&\DGh = i \hbar \spqint \Ki(p)  
\lp \frac{D^2(\th_1)}{16p^2}\rp^{\el}
\Bigg\{
\frac{(-1)^{\de_l} }{\lp 16q^2 \rp^{\abs{\ej}}} 
\lp \G_2^{-1}(q;\L) \,\bG(-q,-p;\L)\rp_{jl} 
-\de_{jl}\, 
\sde(p-q) \Bigg\}
 \nonumber \\
&&\phantom{\DGh = -i\spqint} \!\!\!
\times\lp \G_2^{-1}(p;\L) \, \DD^{-1}(p) \K^{-1}(p)\rp_{li} \,
\frac{\de^2}{\de\Psi_j(q)\, \de \g_i(p)} \lp \G -\zint \gv (\cp-\bcp) \rp\,.
\eeeq
Notice that at $\L=0$ the cutoff ST identity reduces to $\DGb(0)=0$
and, in the UV limit, becomes the usual ST identity \re{ST0}. Moreover we have 
inserted the factor $\hbar$ in \re{dgh} to put into evidence that $\DGh$ 
vanishes at the tree level.

For the sake of future analysis, we introduce the functional 
\beq\nome{Pi}
\Pi[\Psi,\g;\L]=\G[\Psi,\g;\L]- \half (\Psi, \DD^{-1} \Psi)_{\L\L_0}+
\half (\Psi, \DD^{-1} \Psi)_{0\L_0}
\eeq
differing from the cutoff effective action only in the tree-level 
two-point function, in which the IR cutoff has been removed. With
such a definition, in the \UV limit the tree-level contribution to $\Pi(\L)$
coincides with $\sbrs$, whereas at the tree level  $\G_2(\L)$ contains
the IR cutoff (see \re{gamma2}).
In terms of $\Pi$ the functional  $\DGb$ can be rewritten 
as  
$$
\DGb[\Psi,\g;\L]=
-\spint \Kiu(p)\, \lp \frac{D^2(\th_1)}{16\,p^2}
\rp^{\ei} 
\dfud{\Pi'[\Psi,\g;\L]}{\Psi_i(-p)}\, \dfud{\Pi'[\Psi,\g;\L]}{\g_i(p)}\,,
$$
where 
$\Pi'$ is the expression obtained by removing the gauge fixing term in
\re{Pi}. Thus, in the \UV limit, with the help of \re{ST} one has 
\beq\nome{dgb}
\DGb[\Psi,\g;\L]\to\SS_{\Pi'(\L)}\Pi'(\L) \quad\quad{\mbox{for}}\;\;
\L_0\to\infty
\eeq
at any $\L$. The existence of such a limit is guaranteed in
perturbation theory by the UV finiteness of the cutoff effective
action (perturbative renormalizability). In order to show this
property holds also for $\DGh$, it suffices to recognize that the
presence of cutoff functions having almost non-intersecting supports
forces the loop momenta in \re{dgh} to be of the order of $\L$.
Henceforth we will take the \UV limit in $\DG$.  

\subsection{Perturbative solution of $\DG=0$}
The proof of the ST identity \re{DG} in the RG formalism, with
possible anomalies, is based on induction in the loop number and
closely follows that of non-supersymmetric gauge theories discussed in
\cite{noi,MT,BV}.  For the sake of completeness we resume here the key
issues.

One can show \cite{MT} that the evolution of the vertices of $\DG$ at
the loop ${\ell}$ depends on vertices of $\DG$ itself at lower loop
order, so that if $\DG^{(\ell')}=0$ at any loop order $\ell'<\ell$,
then
\beq\nome{chie}
\LdL\DG^{(\ell)}=0\,.
\eeq
Thus one can analyse $\DG$ at an arbitrary value of $\L$. 
There are two natural choices 
corresponding to $\L=0$ and $\L=\L_R$ much bigger than the subtraction
scale $\mu$, \ie $\L_R=\L_0$. With the former the gauge symmetry condition 
fixes the relevant part of the effective action 
in terms of the physical coupling $g(\mu)$ and provides the boundary
conditions of the RG flow, whereas with the latter 
the gauge symmetry condition determines  the cutoff dependent 
bare couplings.
With this choice the implementation of symmetry is simplified
due to the locality~\footnote{Here and in the following
locality means that each term in the expansion of the functionals in
the gauge field $V$ contains only couplings with non-negative
dimension.} of the 
functionals involved. Although the computation of physical 
vertices is generally cumbersome, this second possibility is more 
convenient in the computation of quantities which do not evolve with
the cutoff $\L$, such as the gauge anomaly.
Hence we will adopt the second possibility in the present
paper.

We now discuss the vanishing of $\DG$.
Also  for this functional we define its relevant part, isolating all 
supersymmetric monomials in the fields, sources and their derivatives 
with ghost number one and dimension three. The rest is included in $\DGi$.

At the UV scale $\DG$ is local, or, more precisely, $\DGi(\L_0)={\cal O}(\frac1{\L_0})$,
so that the irrelevant contributions disappear in the \UV limit.  This
can be understood with the following argument.  From \re{dgb},
$\;\DGb(\L_0)$ is manifestly relevant, since $\Pi(\L_0)=\Pir(\L_0)$,
while it is easy to convince oneself that
$\DGh(\L_0)=\DGhr(\L_0)+{\cal O}(\frac1{\L_0})$.  As a matter of fact,
from \re{dgh} one notices that irrelevant terms may arise from
$\bG[\Phi,\g;\L_0]$ and the cutoff functions. At $\L_0$, $\bG$ is
given by either a relevant vertex or a sequence of relevant vertices
joint by propagators with a cutoff function $\Kin(q+P)$, where $P$ is
a combination of external momenta (see \re{gammab}).  Since the
integral is damped by these cutoff functions, only the contributions
with a restricted number of propagators survive in the
\UV limit. One can infer from power counting that they are of the 
relevant type. 
A similar argument holds for the possible irrelevant contributions 
coming from $\Kiu(p)$. Then \re{chie} ensures the locality of
$\DG(\L)$ at any $\L$.

Once the locality of $\DG(\L)$ is shown, the solvability of the
equation $\DG(\L)=0$ can be proven using cohomological methods
\cite{BRS,piguet}.  This is a consequence of the $\L$-independence
of $\DG$ and the solvability of the same equation at $\L=0$, where the
cohomological problem reduces to the standard one.

Henceforth we will consider the first loop, the generalization to
higher loops being straightforward due to the iterative nature of the
solution.
Using \re{dgb}, at  $\L=\L_0$ and at the first loop  \re{DG} reads 
\beq \nome{fintun}
{\cal S}_{\Pi^{(0)}}\,\Pi^{(1)}(\L_0)
\,+\,\DGhr^{(1)}(\L_0)=0\,.
\eeq
This fine-tuning equation allows to fix some of the relevant 
couplings in $\Pi^{(1)}(\L_0)$. As a matter of fact 
the most general functional $\Pi^{(1)}(\L_0)$ can be cast into the form 
\re{gammarel} and  split into two contributions
\beq \nome{pigreco1}
\Pi^{(1)}(\L_0)=\Piinv^{(1)}(\L_0)+\Pit^{(1)}
(\L_0)\, ,
\eeq
where $\Piinv$ contains all the independent monomials which are
invariant, \ie ${\cal S}_{\Pi^{(0)}}\, \Piinv^{(1)}=0$. The explicit
form of $\Piinv^{(1)}$ is obtained from $\sbrs$ in \re{brsac} and 
\re{matter-action} with the replacement 
$$
(V\,,\,\g_i\,,\,\cp\,,\,\bcp\,, \,g\,,\,\phi\,,\,\bphi)\to 
(\sqrt{z_1}\,V\,,\, \sqrt{z_2}\,\g_i\,,\, \sqrt{z_2\,}\cp\,,\,
\sqrt{z_2}\,\bcp\,,\, z_3 g\,,\, \sqrt{z_4}\,\phi\,, \,\sqrt{z_4}\,\bphi)\,.
$$
The remaining monomials contribute to $\Pit$.
Inserting \re{pigreco1} into \re{fintun}, one finds
$$
{\cal S}_{\Pi^{(0)}}\, \Pit^{(1)}(\L_0)\,=-\, \DGh^{(1)}(\L_0)\,,
$$
which yields the couplings in $\Pit^{(1)}$ since $\DGh^{(1)}(\L_0)$
depends only on $\sbrs$.  An explicit calculation shows that the only
divergences are powers of $\L_0$ according to the dimension of the
relative vertex. In particular dimensionless couplings are finite, due
to the presence in \re{dgh} of cutoff functions having almost
non-intersecting supports~\footnote{See \cite{BV} for the explicit
computation of some of these couplings in non-supersymmetric QCD.}.

As to the couplings $z_i(\L_0)$, which are not involved in the
fine-tuning, one is allowed to set them equal to their physical values
at $\L=0$, \ie $z_i(0)=1$. In the standard language this corresponds
to the renormalization prescriptions.

Instead of solving the fine-tuning equation and determine the 
(cutoff-dependent) couplings of the UV action, in the next section
we will deal with the computation of the
gauge anomaly,  which well illustrates how the method works and
meanwhile is a cutoff independent result. At one loop such 
independence is guaranteed by the absence of the anomaly at the tree level
and by the evolution equation \re{chie}.

\section{Gauge anomaly}

For N=1 SYM within the superspace approach it has been demonstrated
\cite{piguet} that the only possible anomaly is the
supersymmetric extension of the standard Adler-Bardeen anomaly
\cite{abj} and its explicit form is given in ref.~\cite{tonin,GGS}.
As well known, its structure is non-polynomial \cite{tonin,FGPS} and
can be expressed as an infinite series in the gauge field $V$. 
In the following we restrict ourselves to the first term
of this expansion, since higher order polynomials can be inferred \cite{FGPS}
using the consistency condition \cite{BRS} which, at this order, 
forces the one-loop anomaly $\AA^{(1)}$ to obey 
${\cal S}_{\Pi^{(0)}}\AA^{(1)}=0$.

In our framework a violation of the ST identity results in the
impossibility of fixing the relevant couplings $\s_i(\L_0)$ in
$\Pi^{(1)}(\L_0)$ in such a way that \re{fintun} is satisfied.  In
other words, this happens when there are relevant monomials in $\DGh$
which are not trivial cocycles of the cohomology of the BRS operator.

As a first step we write $\DGh$ at one loop order. Performing the \UV limit
in \re{dgh} and setting $\L=\L_0$, one has
\beeq&&
\DGh^{(1)}= i \spqint   \Kiu(p) 
\lq\lp \frac{1}{16q^2} \rp^{\abs{\ej}}
\Kin(q) (-)^{\de_i}
\DD_{jk}(q)
\bG_{ki}^{(0)}(-q,-p;\L) - \de_{ij}\de^8(p-q)\rq
\no\\
&& \ph{\DGh^{(1)}= i \spqint   \Kiu(p) }
\times\lp \frac{D^2(\th_1)}{16p^2}\rp^{\ei}
\frac{\de^2}{\de\Psi_j(q)\, \de \g_i(p)} 
\lp \sbrs -\zint \gv (\cp-\bcp) \rp\,.
\eeeq
Then we isolate the matter contribution in $\DGh^{(1)}$ which, depending on
the representation of the matter fields, can  possibly give rise to the 
anomaly
\beeq \nome{primodelta}
\DGh^{(1)}=\DGh^{\mbox{\scriptsize{SYM}}\,(1)}\!\!\!&+&\!\!
i\spqint \Kiu (p) \frac{\Kin(q)}{q^2}\,
\Biggl[
\frac{\de^2 \bG^{(0)}}{\de \phi(-p) \de \bphi(-q)} \, 
\frac{D^2(\th_1)}{16p^2}
\frac{\de^2 \sbrs}{\de \phi(q) \de \g_\phi(p)} \Biggr. 
\\ \nonumber
& & \;\;\;\;\;\;\;\; + \Biggl.\,D\to\bD,\; \phi\to\bphi, \;
\g_\phi\to\g_{\bphi}\;\Biggr]
\,.
\eeeq 
\begin{figure}[htbp]
\epsfysize=6.5cm
\begin{center}
\epsfbox{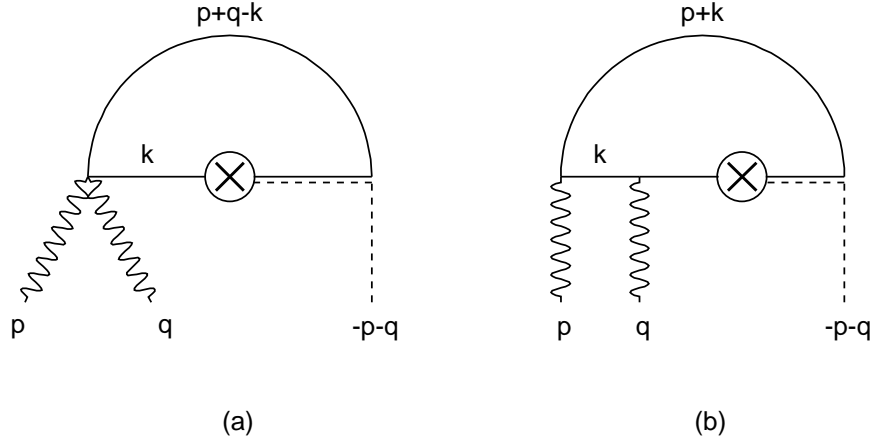}
\end{center}
\caption{\small{Matter contribution to the $\cp$-$V$-$V$ vertex of $\DGh$.
The wavy, dashed and full line denotes the vector, ghost and matter
fields respectively; the double line represents the BRS source
associated to the matter field. The cross denotes
the insertion of the cutoff function $\Kiu$ in the product of the
$\cp$-$\phi$-$\g_\phi$ vertex of $\sbrs$ with: (a) the irreducible
$\bphi$-$V$-$V$-$\phi$ vertex of $\bG$; (b) the reducible
$\bphi$-$V$-$V$-$\phi$ vertex of $\bG$.  All external momenta are
incoming and integration over the loop momentum is understood.  }}
\end{figure}
\newline
Inserting \re{gammab} in \re{primodelta} and extracting the tree-level
vertices of $\bG$ from $\sbrs$, one can see that the matter
contribution to the $\cp$-$V$-$V$ vertex of $\DGh$ is made of two
pieces, as shown in fig.~1.  
The first, originating from the irreducible
part of the $\bphi$-$V$-$V$-$\phi$ vertex of $\bG$, 
is given by
\beeq\nome{cvva}&&
-\frac{ig^2}{32}
\spqint \tr[\cp (-p-q,\th_1) V(p,\th_2) V(q,\th_2)]
\int\frac{d^4k}{(2\pi)^4}
\frac{\Kiu(k)\Kin(p+q-k)}{(p+q-k)^2}\, 
\no\\&&
\phantom{-\frac{ig^2}{32}
\spqint \tr} 
\times \de^4 (\th_1-\th_2) \,\bD^2 D^2 \de^4 (\th_1-\th_2)
\eeeq
and, as suggested from the graph depicted in fig.~1a which is not
typically triangle-shaped, does not contribute to the anomaly.  As a
matter of fact, by restricting to the Yang-Mills sector, one
immediately recognizes that the anti-symmetric tensor $\eps_{\m\n\r\s}$
can not be generated from such a term.  Indeed using \re{susydelta}
and performing the loop integration, the expression in \re{cvva}
becomes
$$
g^2
\int\frac{d^4p\,d^4p\,d^4\th}{(2\pi)^8}\lgr(a_1\,\L_0^2+ a_2\,(p+q)^2)
\tr\lq\cp (-p-q,\th) V(p,\th) V(q,\th)\rq 
+\OO ((p+q)^4/\L_0^2)\rgr 
$$ 
where the $a_i$'s are finite
cutoff-dependent numbers which can be explicitly computed once the
cutoff function is specified.  The finiteness of such coefficients is
due to the presence of cutoff functions having almost non-intersecting
supports, \ie $k^2\lesssim \L_0^2$ and $(p+q-k)^2\gtrsim \L_0^2$.
These two monomials belong to the trivial cohomology of $\SS_\G$ and
their coefficients, together with those stemming from analogous
monomials of $\DGh^{\mbox{\scriptsize{SYM}}}$, fix the parameters in
$\Pit^{(1)}$ via \re{fintun}.

We turn now to the contribution associated to the 
graph represented in fig.~1b, which originates from the second term in
the iterative expansion of $\bG$ in vertices of $\G$.
It reads
\beeq\nome{cvvb}&&
i\frac{g^2}{256}\,
\spqint \int\frac{d^4k}{(2\pi)^4}
\tr[\cp (-p-q,\th_2) V(p,\th_1) V(q,\th_2)]\,
\frac{\Kiu(k-q)\Kin(p+k)\Kin(k)}{k^2(k+p)^2} 
\no\\&&
\phantom{i\frac{g^2}{256}\,
\spqint \int\frac{d^4k}{(2\pi)^4}\tr}
\times
\bD^2 D^2(k,\th_1)\, \de^4 (\th_1-\th_2) \,
D^2 \bD^2(k+p,\th_1)\, \de^4 (\th_1-\th_2)\,.
\eeeq
After integrating the $\bD^2 D^2$ derivatives by parts and using the 
algebra of covariant derivatives (reported in the appendix)  
and \re{susydelta}, one finds that 
the only non-vanishing terms in  \re{cvvb} are
\beeq\nome{cvvb'}
&&
i\frac{g^2}{256}\,
\int\frac{d^4p\, d^4q \,d^4\th}{(2\pi)^8}\int\frac{d^4k}{(2\pi)^4}\;
\frac{\Kiu(k-q)\Kin(k)\Kin(p+k)}{k^2(k+p)^2}
\no\\&&\phantom{i\frac{g^2}{256}\,}
\times
\tr \lq\cp (-p-q,\th)
\lp\lp  \bD^2 D^2+ 8  k_{\a\ad}\bD^\ad D^\a+16 k^2\rp  V(p,\th)\rp\, 
V(q,\th)\rq\,.
\eeeq
By performing the loop integration one finds out that
the first and the third term in the trace  generate
only monomials which belong to the trivial
cohomology of $\SS_\G$, \ie
\beeq&&
g^2\int\frac{d^4p\,d^4q\,d^4\th}{(2\pi)^8}\lgr a_3
\tr\lq\cp (-p-q,\th) \lp\bD^2\,D^2 V(p,\th) \rp\,V(q,\th)\rq\right.
\no\\&&
\phantom{ g^2\int\frac{d^4p\,d^4q\,d^4\th}{(2\pi)^8} } \left.
+ (a_4\,\L_0^2+ a_5\, P^2 )\tr\lq\cp (-p-q,\th) V(p,\th)V(q,\th)\rq
+\OO (P^4/\L_0^2)\rgr\no
\eeeq
where $P$ is some combination of the momenta $p$ and $q$ and the
$a_i$'s are finite cutoff-dependent numbers.
We are now left with the second term in the trace in \re{cvvb'}.
Exploiting symmetry properties and expanding into external momenta we obtain
\beq \nome{anomaly1}
\frac{g^2}{1024\,\pi^2}
\int\frac{d^4p\,d^4q\,d^4\th}{(2\pi)^8}
\tr[ \cp (-p-q,\th) \lp\bD^\ad D^\a V(p,\th)\rp\,V(q,\th)]\,
(q_{\a\ad}\,I_1+  p_{\a\ad}\,I_2)
\eeq
where
\beeq&&
I_1=
\int_0^{\infty} dx \,\Kin^2 (x)\dpad{\Kiu(x)}{x} +\OO(P^2/\L_0^2)\no\\&&
I_2=\int_0^{\infty} dx \lq
\Kin^2(x)\dpad{\Kiu(x)}{x} + \frac{\Kin^2(x)}{x}\Kiu(x)\rq
+\OO(P^2/\L_0^2)\no
\eeeq
with $x=k^2/\L_0^2$ and $P$ as above.  Notice that in the \UV limit
$I_1$ yields a cutoff independent number, \ie $-1/3$, since it is
determined only by the values $\Kin(0)=0$ and $\Kin(\infty)=1$. On the
contrary $I_2$ depends on the choice of the cutoff function.

In \re{anomaly1} the structure proportional to $p_{\a\ad}$ does not
contribute to the anomaly, basically because, in the coordinate space,
all derivatives act on the same superfield.  On the other hand, had it
played a role in determining the anomaly, like all other contributions
analyzed above, our method would have led to an inconsistent result,
as $I_2$ and the $a_i$'s depend on the cutoff function.  Hence, only
the term with $q_{\a\ad}$ can generate a genuine anomaly.  By setting
$I_1=-1/3$ in \re{anomaly1} one gets
\beq \nome{anomaly2}
\frac{g^2}{3072\,\pi^2}
\int\frac{d^4p\, d^4q \,d^4\th}{(2\pi)^8}
\tr[\cp (-p-q,\th) \,
\lp\bD^\ad D^\a V(p,\th)\rp\,q_{\a\ad}\,V(q,\th)]
\eeq
which has the true structure of the anomaly.  

The $\bcp$-$V$-$V$ vertex of $\DGh$ can be derived repeating the
steps described above. Also in this case one identifies the anomalous
contribution by isolating its cutoff independent part, which turns out to be
\beq \nome{anomaly3}
-\frac{g^2}{3072\,\pi^2}
\int\frac{d^4p\, d^4q \,d^4\th}{(2\pi)^8}
\tr[\bcp (-p-q,\th) \,
\lp D^\a \bD^\ad V(p,\th)\rp q_{\a\ad}\,V(q,\th)]\,.
\eeq
Finally, summing up  \re{anomaly2} and \re{anomaly3}, and switching to 
the coordinate space, the anomaly has the well-known form
\beq \nome{anomaly}
\AA=\frac{g^2}{6144\,\pi^2}
\zint\lp
\tr[\cp \,\bD^\ad D^\a V\,\{D_\a,\bD_{\ad}\}V]
-\tr[\bcp\,D^\a \bD^\ad V \,\{D_\a,\bD_{\ad}\}V]\rp\,.
\eeq
As a remark, we notice that in order to reproduce the standard abelian
anomaly in non-supersymmetric QCD one has to perform the integration
over the grassmannian variables, identify the ghost $c$ with
$\cp+\bcp$ and replace $g$ with $2g$ to recover the usual
gluon-fermion coupling (see eq.~\re{matter-action}).  Then one finds
that the coefficient of the monomial $\eps^{\m\n\r\s}\tr[\p_\m c\, \p_\r
A_\n A_\s]$ is exacly $g^2/(24\pi^2)$.

\section{Conclusions}

In this paper we considered supersymmetric (gauge) theories within the
RG approach. Although we restricted to the WZ model and N=1 SYM, the
formalism is developed in such a way it can be applied to any
supersymmetric theory with an arbitrary field content and with
extended supersymmetry.

An advantage of the RG formulation is that the regularization is
implemented by introducing a cutoff in the loop momenta which makes all
the Green functions UV finite. This means one need not analitycally
continue the Feynman integrals in the space-time dimension $d$, which
is kept fixed (in our case $d=4$). Therefore both the equality of
bosonic and fermionic degrees of freedom is safe --a necessary
condition for supersymmetry-- and the superspace technique presents no
ambiguity, for instance in handling the algebra of covariant
derivatives, traces of $\s$ matrices and using Fierz identities.

However, in the RG approach the presence of the cutoff explicitly
breaks gauge symmetry. This is an unavoidable consequence of the
absence of a regularization scheme that manifestly preserves both
supersymmetry and BRS invariance, which in turn is intimately related
to the existence of the chiral anomaly.

In this paper we showed that the Slavnov-Taylor identity for the
physical effective action of an anomaly-free theory
is perturbatively recovered by solving the fine-tuning equation
\re{DG} at the UV scale. Such a procedure was sketched in subsect.~4.1. 
On the other hand, in case of unfulfilled matching conditions for the anomaly
cancellation,
we reproduced the supersymmetric 
chiral anomaly by a simple one-loop
calculation.
We performed a one-loop analysis, but the procedure systematically
generalizes to higher order.

In our framework one can derive the non-renormalization theorem for the
WZ model with no substantial modification with respect to the standard
proof \cite{Grisaru}. In the massless case, it is straightforward 
recognizing that chiral superfield
interactions of the type $\cint \lp \z\,\phi + m\,\phi^2+\l\,\phi^3
\rp $ do not
receive any finite or infinite perturbative contributions at the first
loop.  As a matter of fact, inserting \re{gammab} at the tree level in
\re{eveq} one can see that only vertices with an equal number of
chiral and anti-chiral fields acquire one-loop corrections. At higher
loops the mechanism is less obvious, since there are reducible
vertices of $\bG_{\phi\bphi}$ with any number of $\phi$ and $\bphi$,
which apparently give rise to interactions like $\phi^m\bphi^n$, even
with $n\ne m$.  However, by repeatedly using the rules of covariant
derivative algebra, one is easily convinced that all the interactions
with $n\ne m$ vanish.  More generally, this argument can be extended
to the massive case.

As well known, in the superspace formulation of SYM 
one has to face with the problem of infrared singularities,
due to the appearance of the pseudoscalar field $C(x)$, the $
\theta = 0 $ component of the gauge superfield (this difficulty is 
obviously circumvented in the Wess-Zumino gauge
\cite{wzgauge}, where the field $C$ is absent).
To avoid this problem one can assume \cite{piglect} that all fields are
made massive by adding suitable supersymmetric mass terms in the
action.  Since these masses break BRS invariance, the corresponding
Slavnov-Taylor identity holds only in the asymptotic region of
momentum space.

In our formulation the presence of the IR cutoff $\L$ naturally makes
all cutoff vertices IR finite for $\L\ne 0$. Furthermore, for a
non-supersymmetric massless theory it has been proven, by induction in
the number of loops \cite{IR}, that the vertex functions without
exceptional momenta are finite for $\L\to 0$, once the relevant
couplings are defined in terms of cutoff vertices evaluated at some
non-vanishing subtraction points. In this proof the convergence of
loop integrals for $\L\to 0$ is simply controlled by the number of
soft momenta in the vertices which appear in the iterative solution of
the RG equation \re{eveq}. Therefore we believe its generalization to
the supersymmetric case presents no difficulty.

Finally, though we restricted our analysis to the perturbative regime,
the RG formulation is in principle non-perturbative and provides a
natural context in which to clarify the connection between exact
results and those obtained in perturbation theory. In particular, it
would be interesting to consider issues such as the anomaly puzzle and
the violation of holomorphicity
\cite{shifman}.

\vskip1truecm 
\noindent
{\it Note added:} After the completion of this paper, a paper by
S. Falkenberg and B. Geyer \cite{falk} appeared in which the RG
formalism in the effective average action approach \cite{Wett} is
applied to N=1 SYM. In this paper an approximate solution of the RG
flow for SYM within the background field method is obtained by a
truncation of the average action. Nevertheless, as remarked by the
authors themselves, such a truncation conflicts with BRS invariance.

\vspace{1cm}\noindent{\large\bf Acknowledgements}
We wish to thank K. Lechner and  M. Tonin for useful discussions.

\section*{Appendix: supersymmetric conventions}

The notations and conventions are those of~\cite{piglect}. 
Given a  Weyl spinor  $\psi_\a$,  $\a =1,2$,  indices can be raised and lowered
as follows
$$
\psi^\a=\eps^{\a\b}\psi_\b\, ,\quad \psi_\a=\eps_{\a\b}\psi^\b\, ,
$$
with  
$$
\eps^{\a\b}=-\eps^{\b\a}\, ,\quad\eps^{12}=1\, ,\quad
  \eps_{\a\b}=-\eps^{a\b}\, ,\quad \eps^{\a\b}\eps_{\b\g}=\de^\a_\g\, ,
$$
(the same for dotted indices). The 
summation convention is $\psi\chi=\psi^\a\chi_\a$ and 
$\bpsi\bchi=\bpsi_\ad\bchi^\ad$.

The matrices $\s^\m$ with lower indices are
$$
\s^\m_{\a\bd}=(\identity, \s^i)_{\a\bd}\,,
$$ 
where the $\s^i$'s are the Pauli matrices, whereas those with 
upper indices are
$$
\bs_\m^{\ad\b}=\s_\m^{\b\ad}
    =\eps^{\b\a}\eps^{\ad\bd}\s_{\m\,\a\bd}\,.
$$
A vector superfield $V(x,\th,\bt)$ has the following expansion
\beeq&&
V(x,\th,\bt) = C(x) + \th\chi(x) + \bt\bar\chi(x) + \half\th^2M(x)
+ \half\bt^2\bar M(x)  \no\\&&
\phantom{V(x,\th,\bt)}
+ \th\s^\m\bt A_\m(x) + \half\bt^2\th\l(x)
+ \half\th^2\bt\bar\l + \frac 1 4  \th^2\bt^2D(x)\, ,
\eeeq
where the components are ordinary space-time fields. 
A chiral (anti-chiral) superfield $\phi$  ($\bphi$)
expanded in component fields is
\beeq&&
\phi(x,\th,\bt) = 
  e^{-i\th\s^\m\bt\p_\m}\lp \phi(x) + \th\psi(x) + \th^2 F(x)\rp\,
\no\\&&
\bphi(x,\th,\bt) =
e^{i\th\s^\m\bt\p_\m}\lp \bar \phi(x) + \bt\bpsi(x) + \bt^2 \bar F(x)\rp\,.
\eeeq
The components of a vector superfield
transform under supersymmetry as
\beq\begin{array}{ll}
\de_\a C = \chi                
            &\bar\de_\ad C = \bar\chi \\[2mm]
\de_\a \chi^\b = \de_\a^\b M   
            &\bar\de_\ad {\bar\chi}^\bd=-\de_\ad^\bd\bar M\\[2mm]
\de_\a\bar\xi_\ad = \smuaad (A_\m+i\p_\m C) 
            &\bar\de_\ad\xi_\a = -\smuaad (A_\m-i\p_\m C) \\[2mm]
\de_\a M = 0  
            &\bar\de_\ad \bar M = 0 \\[2mm]
\de_\a \bar M = \l_\a-i(\s^\m\p_\m\bar\chi)_\a
            &\bar\de_\ad  M = \bar\l_\ad+i(\p_\m\chi\s^\m)_\ad \\[2mm]
\de_\a A_\m = \half(\s_\m\bar\l)_\a - \frac{i}{2}(\s^\n\bs_\m\p_\n\chi)_\a\quad
            &\bar\de_\ad A_\m = \half(\l\s_\m)_\ad 
                     + \frac{i}{2}(\p_\n\bar\chi\bs_\m\s^\n)_\ad \\[2mm]
\de_\a\l^\b = \de_\a^\b D + i(\s^\n\bs^\m)_\a{}^\b\p_\n A_\m
            &\bar\de_\ad{\bar\l}^\bd = -\de_\ad^\bd D 
                     + i(\bs^\m\s^\n)^\bd{}_\ad \p_\n A_\m \\[2mm]
\de_\a\bar\l_\ad = i\smuaad\p_\m M  
            &\bar\de_\ad\l_\a = i\smuaad\p_\m\bar M  \\[2mm]
\de_\a D = -i(\s^\m\p_\m\bar\l)_\a
            &\bar\de_\ad D = i(\p_\m\l\s^\m)_\ad\,.
\end{array}
\eeq
For the components of the chiral and anti-chiral superfields one has
\beq\begin{array}{ll}
\de_\a \phi = \psi_\a          &\bar\de_\ad \bphi = \bpsi_\ad   \\[2mm]
\de_\a\psi^\b = 2\de_\a^\b F   &\bar\de_\ad\bpsi^\bd = -2\de_\ad^\bd\bar F \\[2mm]
\de_\a F = 0                 &\bar\de_\ad\bar F = 0  \\[2mm]
\de_\a\bphi = 0              &\bar\de_\ad \phi= 0  \\[2mm]
\de_\a\bpsi_\ad = 2i\smuaad\p_\m\bphi \quad
                             &\bar\de_\ad\p_\a = 2i\smuaad\p_\m \phi  \\[2mm]
\de_\a\bar F = -i(\s^\m\p_\m\bpsi)_\a \quad
                             &\bar\de_\ad F = i(\p_\m\p\s^\m)_\ad \,.
\end{array}
\eeq
The covariant derivatives,  defined 
such as to anti-commute with the supersymmetry
transformation rules,  are given by
\beq
D_\a  =  \dpad{}{\th^\a} - i\smuaad\bt^\ad\p_\m \, ,\quad
\bD_\ad =  -\dpad{}{\bt^\ad} + i\th^\a\smuaad\p_\m \, .
\eeq
They obey the algebra
\beq
\lgr D_\a,\bD_\ad \rgr = 2i\smuaad\p_\m
\eeq
(the other anti-commutators vanish). Useful relations these covariant
derivatives satisfy are
\beeq&&
[D_\a,\bD^2]=4i(\s^\m\bD)_\a\p_\m\, ,\quad
[\bD_\ad,D^2]=-4i(D\s^\m)_\ad\p_\m \no\\&&
[D^2,\bD^2]=8iD\s^\m\bD\p_\m+16\p^2 = 
           -8i\bD\bs^\m D\p_\m-16\p^2\no\\&&
D\bD^2D=\bD D^2\bD\no\\&&
D\bD_\ad D = -\half\bD_\ad D^2-\half  D^2\bD_\ad\, , \quad
\bD D_\a \bD = -\half D_\a \bD^2-\half  \bD^2D_\a\,.
\eeeq

The superspace integral
of a superfield $V$, or of a (anti)chiral superfield 
$\phi$ ($\bar \phi$) is given by
\beq
\zint V=\int d^4x\, D^2\,\bD^2 V\,,\quad
\cint  \phi=\int   d^4x\, D^2 \phi\,,\quad
\acint \bar \phi=\int d^4x\, \bD^2\bphi\,,
\eeq
the integral with respect to the Grassmann variable $\th$
being defined by the derivative $\p/\p\th$.

The following  operators
\beq
P^{\rm T} = \frac{D\bD^2D}{8\p^2}\, , \quad
P^{\rm L} = -\frac{D^2\bD^2+\bD^2D^2}{16\p^2} 
\eeq
are projectors. In particular, $P^{\rm L}$ can be used to write
integrals of chiral (or anti-chiral) superfields as integrals over the
full superspace measure (recall that only for this measure
the integration by parts holds). For instance 
$\int\frac{d^4p\,d^2\th}{(2\pi)^4}\, \phi= \spint \frac{D^2}{16p^2}\,\phi$.

The delta function is defined by
$$
\de^8(z_1-z_2) = \de^4(\th_1-\th_2) \,\de^4(x_1-x_2) =
\frac{1}{16}(\th_1-\th_2)^2\,(\bt_1-\bt_2)^2
\,\de^4(x_1-x_2)\,.
$$
The functional derivatives are 
\beq
\dfud{V(z_1)}{V(z_2)}=\de^8(z_1-z_2)\,,
\quad
\dfud{\phi(z_1)}{\phi(z_2)}=\bD^2\de^8(z_1-z_2)\,,\quad
\dfud{\bphi(z_1)}{\bphi(z_2)}=D^2\de^8(z_1-z_2)\,.
\eeq

Finally, in order to separate the trivial cocycles from the anomaly in
\re{primodelta}, it can be useful to switch to components. Then for
the non-supersymmetric YM sector the anomaly is proportional to
$\eps_{\m\n\r\s}$, which is generated by the following trace
$$
\tr\lq
\s^\m\bs^\n\s^\r\bs^\ta\rq = 2\lp g^{\m\n}g^{\r\ta} + g^{\n\r}g^{\m\ta} -
 g^{\m\r}g^{\n\ta}
  - i\eps^{\m\n\r\ta}\rp \,.
$$

\end{document}